\documentclass[12pt]{iopart}

\usepackage{amssymb,amsthm}
\usepackage{graphicx,cite}

\begin{document}

\title[The Shannon-entropy uncertainty relation for $D$-dimensional central potentials]{The Shannon-entropy-based uncertainty relation for $D$-dimensional central potentials}

\author{\L ukasz Rudnicki$^1$, Pablo S\'anchez-Moreno$^{2,3}$ and Jes\'us S. Dehesa$^{3,4}$}

\address{$^1$ Center for Theoretical Physics, Polish Academy of Sciences, Warsaw, Poland}
\address{$^2$ Department of Applied Mathematics, University of Granada, Granada, Spain}
\address{$^3$ Institute Carlos I for Computational and Theoretical Physics, University of Granada, Granada, Spain}
\address{$^4$ Department of Atomic, Molecular and Nuclear Physics, University of Granada, Granada, Spain}

\ead{rudnicki@cft.edu.pl,pablos@ugr.es,dehesa@ugr.es}

\begin{abstract}
The uncertainty relation based on the Shannon entropies of the probability densities in position and momentum spaces is improved for quantum systems in arbitrary $D$-dimensional spherically symmetric potentials. To find it, we have used the $L^p$ -- $L^q$ norm inequality of L. De Carli
and the logarithmic uncertainty relation for the Hankel transform of S. Omri.
Applications to some relevant three-dimensional central potentials are shown.
\end{abstract}


%


\section{Introduction}

The position-momentum uncertainty principle is likely the most prominent difference between classical and quantum physics. The first mathematical realization of this principle was done by W. Heisenberg \cite{heisenberg_zp27} and E. Kennard \cite{kennard_zp27} in terms of the standard deviations of the quantum-mechanical probability densities of the particle in position and momentum spaces. However the so-called Heisenberg uncertainty relation is neither the most appropriate nor the most stringent \cite{hilgevoord_ejp85,majernik_ejp97,uffink_thesis}. Indeed, the standard deviation is a measure of separation of the region(s) of concentration of the probability cloud from the centroid (a particular point of the distribution), rather than a measure of the extent to which the distribution is in fact concentrated \cite{hall_pra99,majernik_ejp97,uffink_thesis}. Information theory \cite{cover_91,frieden_04} provides more appropriate local (Fisher information) and global (Shannon, R\'enyi and Tsallis entropies) uncertainty measures. The Shannon entropy has been argued to be the best global measure of the spreading of a probability distribution according to some criteria \cite{shannon:bst48,cover_91,uffink_thesis}; see also \cite{hatori_kmsr58,csiszar_e08} for further details.

The (differential) Shannon entropies of the position and momentum continuous probability distributions, $\rho(\vec{r})$ and $\gamma(\vec{p})$, respectively, are known to fulfil the so-called entropic uncertainty relation (units with $\hbar=1$ are used)
\begin{equation}
S[\rho] + S[\gamma] \ge D(1+\ln\pi)
\label{eq:uncertainty_relation_bbm}
\end{equation}
where
\begin{equation}
S[\rho]= -\langle\ln\rho\rangle = -\int_{\mathbb{R}^D} \rho(\vec{r})\ln\rho(\vec{r})d^Dr
\label{eq:shannon_definition}
\end{equation}
denotes the $D$-dimensional position Shannon entropy, and $S[\gamma]$ denotes the corresponding momentum Shannon entropy. This relation was conjectured independently in 1957 by H. Everett \cite{everett_73} and I.I. Hirschman \cite{hirschman_amj57} and proved in 1975 by W. Beckner \cite{beckner_am75} and I. Bialynicki-Birula and J. Mycielski \cite{bialynicki_cmp75}. It provides a strict improvement upon the standard Heisenberg relation \cite{bialynicki_cmp75,uffink_thesis}.

The R\'enyi entropy of the position probability $\rho(\vec{r})$ is defined \cite{renyi_70} by
\begin{equation}
R_q[\rho]=\frac{1}{1-q}\ln\int_{\mathbb{R}^D} \left[\rho(\vec{r})\right]^q d^Dr,	\quad q>0, q\neq 1.
\label{eq:renyi_difinition}
\end{equation}
Notice that this quantity is a generalization of the Shannon entropy given by (\ref{eq:shannon_definition}) so that
\[
\lim_{q\to 1} R_q[\rho]=S[\rho].
\]
Due to this property it is often convenient to consider the general case of the R\'enyi entropy and use the limit $q\to 1$ obtain the particular results for the Shannon entropy. We shall use this strategy in Section \ref{sec:radial_wavefunctions}.

The main goal of this paper is to refine the entropic uncertainty relation (\ref{eq:uncertainty_relation_bbm}) for particles moving in a central potential $V_D(r)$, $r=|\vec{r}|$.
The $D$-dimensional ($D \ge 2$) central-field approximation has been successfully applied in one- and many-body physics from the early days of quantum mechanics  up until now, to explain numerous physical properties and phenomena of natural systems (see e.g., \cite{bohr_zp22,fock_zp30,hartree_pcps28,kitagawara_jpb83,chatterjee_pr90,herschbach_93,anglin_n02,wesson_06}). For example, it is known to be the theoretical basis of the periodic table of the chemical elements \cite{mendeleev_jrpcs69} together with the Pauli exclusion principle. Moreover, the central potentials are being very often used as prototypes for numerous other purposes and systems not only in the three-dimensional world (e.g., oscillator-, Coulomb-, and van der Waals-like potentials) but also in non-relativistic and relativistic $D$-dimensional physics (see e.g., \cite{chatterjee_pr90,herschbach_93,wesson_06}). Recently, it has been also applied to study the behaviour of the quantum dots and wires as well as to interpret the experiments of dilute bosonic and fermionic systems in magnetic traps at extremely low temperatures \cite{gleisberg_pra00,anglin_n02,demarco_s99}, what is provoking a fast development of a density-functional theory of independent particles moving in multidimensional central potentials \cite{howard:pra02,dehesa_njp07}.

We begin with the Schr\"odinger equation of the corresponding $D$-dimensional central force problem:
\begin{equation}
\left[ -\frac12 \vec{\nabla}_D^2 + V_D(r)\right] \Psi(\vec{r}) = E \Psi(\vec{r}),
\label{eq:schrodinger_equation}
\end{equation}
$\vec{\nabla}_D^2$ denoting the Laplace operator \cite{chatterjee_pr90,avery_00,romera_jmp06} associated with the position vector in hyperspherical coordinates $\vec{r}=(r,\theta_1,\ldots,\theta_{D-1})\equiv(r,\Omega_{D-1})$, $\Omega_{D-1}\in S^{D-1}$ (surface of the $D$-dimensional sphere), which is given by
\[
\vec{\nabla}^2_D \equiv \frac{\partial^2}{\partial r^2} + \frac{D-1}{r}\frac{\partial}{\partial r}
-\frac{\Lambda^2_{D-1}}{r^2}
\]
and the squared hyperangular momentum operator $\Lambda_{D-1}^2$ is known to fulfil the eigenvalue equation
\[
\Lambda^2_{D-1} \mathcal{Y}_{l,\{\mu\}}(\Omega_{D-1})=l(l+D-2)\mathcal{Y}_{l,\{\mu\}}(\Omega_{D-1}).
\]
The symbol $\mathcal{Y}_{l,\{\mu\}} (\Omega_{D-1})$ denotes the hyperspherical harmonics \cite{avery_00,dehesa_ijqc10} characterized by the $D-1$ hyperangular quantum numbers $(l\equiv\mu_1,\mu_2,\ldots,\mu_{D-1}\equiv m)\equiv (l,\{\mu\})$, which are integer numbers with values $l=0,1,2,\ldots$, and $l\ge\mu_2\ge\ldots\ge \mu_{D-2}\ge|\mu_{D-1}|\ge 0$.
Notice that for $D=2$ we only have one quantum number $l\in\mathbb{Z}$.

The eigenfunctions $\Psi(\vec{r})$ of this problem may be written as
\begin{equation}
\Psi(\vec{r})\equiv R(r)\mathcal{Y}_{l,\{\mu\}}(\Omega_{D-1})
=  r^\frac{1-D}{2}u(r)\mathcal{Y}_{l,\{\mu\}}(\Omega_{D-1}),
\label{eq:wavefunction_separation}
\end{equation}
where the reduced radial eigenfunction $u(r)$ is the physical solution of the one-dimensional Schr\"odinger equation in the radial coordinate $r$; namely
\begin{equation}
\left[
-\frac12 \frac{d^2}{dr^2}
+\frac{L(L+1)}{2r^2}+V_D(r)
\right] u_{E,l}(r)
=
E u_{E,l}(r),
\label{eq:radial_schrodinger}
\end{equation}
where we have used the notation $L=l+(D-3)/2$ for the grand hyperangular momentum.
Since $L(L+1)=l(l+D-2)+(D-1)(D-3)/4$, it is worth to realize by looking at Eq. (\ref{eq:radial_schrodinger}) that a particle moving in a $D$-dimensional potential is subject to two additional forces besides the force coming from the external potential $V_D(r)$: the centrifugal force associated with non-vanishing hyperangular momentum, and a quantum fictitious force associated to the so-called quantum-centrifugal potential $(D-1)(D-3)/(8r^2)$ which has a purely dimensional origin \cite{schleich_pra02}. This potential vanishes for $D=1$ and $3$, being negative for $D=2$ and positive for $D\ge 4$. Then the quantum fictitious force, which exists irrespective of the hyperangular momentum and has a quadratic dependence on the dimensionality, has an attractive character when $D=2$ and is repulsive for $D\ge 4$.
Let us also remark that $L=l$ when $D=3$.

From Eq. (\ref{eq:wavefunction_separation}), one can find that the allowed quantum-mechanical state $(E,l,\{\mu\})$ has the following probability density
\begin{equation}
\rho(\vec{r})=\frac{|u(r)|^2}{r^{D-1}} \left| \mathcal{Y}_{l,\{\mu\}}(\Omega_{D-1})\right|^2,
\label{eq:density_position_separation}
\end{equation}
where, according to (\ref{eq:schrodinger_equation}) and (\ref{eq:radial_schrodinger}), we know that $u(r)=u_{E,l}(r)$ and $\rho(\vec{r})=\rho_{E,l,\{\mu\}}(\vec{r})$. In addition, the normalization-to-unity of the wave function $\Psi(\vec{r})\equiv \Psi_{E,l,\{\mu\}}(\vec{r})$ yields that
\[
\int_0^\infty |u(r)|^2 dr = 1
\]
for the reduced radial eigenfunction, where we have taken into account the normalization condition of the hyperspherical harmonics:
\[
\int_{S^{D-1}} \left|\mathcal{Y}_{l,\{\mu\}}(\Omega_{D-1})\right|^2 d\Omega_{D-1} =1.
\]

Then, according to Eqs. (\ref{eq:shannon_definition}) and (\ref{eq:density_position_separation}), one has that the Shannon entropy of the $D$-dimensional density $\rho(\vec{r})$ is given by
\begin{equation}
S[\rho] = S[\omega]+(D-1)\langle\ln r\rangle + S(\mathcal{Y}_{l,\{\mu\}}),
\label{eq:shannon_position_separation}
\end{equation}
where $S[\omega]$ denotes the Shannon entropy of the one-dimensional probability density $\omega(r)=|u(r)|^2$, i.e.
\[
S[\omega]=-\int_0^\infty \omega(r)\ln \omega(r)dr.
\]
Similarly, $S(\mathcal{Y}_{l,\{\mu\}})$ gives the Shannon entropy of the hyperangular probability density $|\mathcal{Y}_{l,\{\mu\}}|^2$, i.e.
\begin{equation}
S(\mathcal{Y}_{l,\{\mu\}}) = -\int_{S^{D-1}} \left| \mathcal{Y}_{l,\{\mu\}} \right|^2 \ln\left|\mathcal{Y}_{l,\{\mu\}}\right|^2 d\Omega_{D-1}
\label{eq:entropy_y}
\end{equation}
with the volume element
\[
d\Omega_{D-1} = \prod_{j=1}^{D-1}(\sin\theta_j)^{\alpha_j}d\theta_j,\quad \alpha_j=D-j-1.
\]
Moreover, the logarithmic expectation value $\langle\ln r\rangle$ is defined as
\[
\langle \ln r\rangle =\int_{\mathbb{R}^D}\rho(\vec{r}) \ln r \: d^D r
=\int_0^\infty \omega(r) \ln r \: dr
\]

A parallel analysis in momentum space yields the following expression
\begin{equation}
S[\gamma] = S[\tilde{\omega}]+(D-1)\langle \ln p \rangle + S(\mathcal{Y}_{l,\{\mu\}}),
\label{eq:shannon_momentum_separation}
\end{equation}
for the momentum Shannon entropy of a particle in the $D$-dimensional central potential $V_D(r)$, where $S[\tilde{\omega}]$ and $\langle\ln p \rangle$ have analogous expressions to $S[\omega]$ and $\langle\ln r\rangle$ but with respect to the momentum density
\[
\gamma(\vec{p}) = \left|\tilde{\Psi}(\vec{p})\right|^2 = \frac{\left|\tilde{u}(p)\right|^2}{p^{D-1}}\left|\mathcal{Y}_{l,\{\mu\}}(\Omega_{D-1})\right|^2
\]
where $\tilde{\Psi}(\vec{p})$ is the Fourier transform of $\Psi(\vec{r})$, (keep in mind that the hyperspherical harmonics remains invariant under this transformation). The reduced radial momentum eigenfunction $\tilde{u}(p)$ is related to $u(r)$ by means of the Hankel transform
\begin{equation}
\tilde{u}(p) =(-i)^l \int_0^\infty \sqrt{rp} J_{l+D/2-1}(rp) u(r) dr,
\label{eq:hankel_transform}
\end{equation}
the $J_\nu(x)$-symbol being the first-kind Bessel function of order $\nu$. Let us call $\tilde{\omega}(p)=|\tilde{u}(p)|^2$ the momentum reduced radial probability density. The overall phase $\left(-i\right)^l$ will play no role in our investigations.

Summing Eqs. (\ref{eq:shannon_position_separation}) and (\ref{eq:shannon_momentum_separation}) we obtain:
\begin{equation}
S[\rho] + S[\gamma] = S[\omega]+S[\tilde{\omega}] + (D-1)(\langle \ln r\rangle + \langle \ln p\rangle) + 2 S(\mathcal{Y}_{l,\{\mu\}}),
\label{eq:entropy_sum_expression}
\end{equation}
which will be the basic starting point to obtain our goal. Let us emphasize that the hyperangular Shannon entropy $S\left(\mathcal{Y}_{l,\{\mu\}} \right)$ is under control since the hyperspherical harmonics are well-known mathematical objects so that they do not depend on the external potential $V_D(r)$. On the other hand, the remaining terms $S[\omega]+S[\tilde{\omega}]$ and $\langle\ln r\rangle + \langle \ln p\rangle$ do depend on $V_D(r)$; so that, we will try to bound them from below in terms of the hyperquantum number $l$ which characterizes the state. This will be done in Sections \ref{sec:radial_wavefunctions} and \ref{sec:logarithmic} by use of the $L^p$ -- $L^q$ norm inequality of L. De Carli \cite{decarli_jmaa08} and the logarithmic uncertainty relation of S. Omri \cite{omri_itsf11}, respectively. In Section \ref{sec:uncertainty_relation}, the new entropic uncertainty relation is given and discussed.  Let us advance that the encountered lower bound to the Shannon-entropy sum only depends on the  hyperangular quantum numbers in an analytical form. Then, in Section \ref{sec:applications} the new entropic relation is examined for two relevant three-dimensional central potentials: the Coulomb and the harmonic oscillator potentials. Finally, some conclusions and open problems are given in Section \ref{sec:conclusions}.

\section{Entropic uncertainty relations for the reduced radial wavefunctions}
\label{sec:radial_wavefunctions}

In this Section we will find the R\'enyi-entropy-based and the Shannon-entropy-based uncertainty relations of the reduced radial wave functions $u(r)$ and $\tilde{u}(p)$ in position and momentum spaces respectively.

Taking into account that $\tilde{u}(p)$ is the Hankel transform (\ref{eq:hankel_transform}) of $u(r)$, we can directly apply the Theorem 1.2 of L. De Carli \cite{decarli_jmaa08} to write that
\begin{equation}
\left( \int_0^\infty \left|\tilde{u}(p)\right|^q dp \right)^\frac{1}{q}
\le C(q,q';\nu) \left(\int_0^\infty  \left|u(r)\right|^{q'} dr\right)^\frac{1}{q'}
\label{eq:decarli_relation}
\end{equation}
for $1<q'\le 2$, and $1/q'+1/q=1$, with the constant
\[
C(q,q';\nu) = \frac{A(q';\nu)}{A(q;\nu)},\quad A(q;\nu)= 2^{\frac{1}{2q}}\frac{q^{\frac12\left(\nu+\frac12+\frac{1}{q}\right)}}{\Gamma\left(\frac{q}{2}\left(\nu+\frac12\right)+\frac12\right)^\frac{1}{q}}
\]
where $\nu=l+D/2-1$.

Let us now rewrite this inequality for the reduced radial probability densities $\omega(r)$ and $\tilde{\omega}(p)$. For this purpose it is convenient to use the parameters $(\alpha,\beta)$ related to $(q,q')$ by $q=2\alpha$ and $q'=2\beta$, so that $1/\alpha+1/\beta=2$. Then Eq. (\ref{eq:decarli_relation}) transforms into the following inequality
\[
\left( \int_0^\infty [\tilde{\omega}(p)]^\alpha dp \right)^\frac{1}{2\alpha}
\le C
\left( \int_0^\infty [\omega(r)]^\beta dr \right)^\frac{1}{2\beta},
\]
where $C=C(2\alpha,2\beta;\nu)$. Then, the Neperian logarithm of this inequality multiplied by the negative factor $1/(1-\alpha)$ (since $q\ge 2 \Leftrightarrow \alpha\ge 1$) yields that
\[
\frac{1}{1-\alpha}\ln \left(\int_0^\infty [\tilde{\omega}(p)]^\alpha dp\right)
\ge
\frac{2\alpha\ln C}{1-\alpha}
-
\frac{1}{1-\beta}\ln \left(\int_0^\infty [\omega(r)]^\beta dr\right),
\]
where we have used the fact that $\alpha/(1-\alpha)=-\beta/(1-\beta)$.

Then, recalling that the $q$th-order R\'enyi entropy of a probability density $f(x)$, with $0<x<\infty$, is given according to Eq. (\ref{eq:renyi_difinition}) by
\[
R_q[f]\equiv \frac{1}{1-q}\ln\int_0^\infty [f(x)]^q dx,
\]
the last inequality gives rise to the following R\'enyi-entropy-based uncertainty relation for the position and momentum reduced radial probability densities:
\begin{equation}
R_\beta[\omega] + R_\alpha[\tilde{\omega}] \ge \frac{2\alpha \ln \left[A(2\alpha;\nu)\right]}{\alpha-1}+\frac{2\beta \ln \left[A(2\beta;\nu)\right]}{\beta-1}.
\label{eq:renyi_uncertainty_w}
\end{equation}
Finally, let us highlight that since $\beta=\alpha/(2\alpha-1)$ and making the limit $\alpha\to 1$ this inequality yields the Shannon-entropy-based uncertainty relation for the reduced radial probability densities $\omega(r)$ and $\tilde{\omega}(p)$:
\begin{equation}
S[\omega]+S[\tilde{\omega}] \ge C'_\nu
\label{eq:shannon_uncertainty_w}
\end{equation}
with
\[
C'_\nu=2l+D+2\ln\left[\frac{\Gamma\left(l+\frac{D}{2}\right)}{2}\right]-(2l+D-1)\psi\left(l+\frac{D}{2}\right)
\]
where
$\psi(x)=\Gamma'(x)/\Gamma(x)$ denotes the well-known digamma function. To obtain (\ref{eq:shannon_uncertainty_w}) from Eq. (\ref{eq:renyi_uncertainty_w}) we have taken into account that the Shannon entropy of a probability density $f(x)$,
$
S[f]=-\int_0^\infty f(x) \ln f(x) dx
$,
is \cite{cover_91} the limiting case $\alpha\to 1$ of the R\'enyi entropy $R_\alpha[f]$.

\section{Logarithmic uncertainty relation for central potentials}
\label{sec:logarithmic}

In this Section we will show that the position and momentum probability densities of a particle moving in a spherically symmetric potential $V_D(r)$ satisfy the following uncertainty relation
\begin{equation}
\langle \ln r \rangle + \langle \ln p \rangle \ge \psi\left(\frac{2l+D}{4}\right) +\ln 2;
\quad l=0,1,2,\ldots
\label{eq:logarithmic_relation}
\end{equation}
This inequality improves for central potentials the Beckner's logarithmic uncertainty relation \cite{beckner_pams95} of general validity, in which the lower bound on the logarithmic sum is $\psi\left(D/4\right)+\ln 2$.

To prove the expression (\ref{eq:logarithmic_relation}) we begin with the 2011-dated logarithmic uncertainty relation of S. Omri \cite{omri_itsf11}
\begin{equation}
\fl \int_0^\infty |f(r)|^2\ln r \:d\lambda_\mu(r) +
\int_0^\infty |\tilde{f}(p)|^2\ln p \:d\lambda_\mu(p) \ge \left[\psi\left(\frac{\mu+1}{2}\right)+\ln 2\right]N_\mu,
\label{eq:omri_relation}
\end{equation}
where $f\in L^{2}\left(0,\infty\right)$, the measure $d\lambda_\mu(r)$, $r\in[0,\infty)$ is given by
\[
d\lambda_\mu (r)= \frac{r^{2\mu+1}}{2^\mu \Gamma(\mu+1)}dr,
\]
and the Hankel transform $\tilde{f}(p)$ of order $\mu$ is defined by
\[
\tilde{f}(p)=\int_0^\infty f(r) j_\mu(rp)d\lambda_\mu(r), \quad  \mu\ge-\frac12,
\]
where $j_\mu(r)$ is the normalized spherical Bessel function of the first kind, given by
\[
j_\mu(z)=\frac{2^\mu \Gamma(\mu+1)}{z^\mu}J_\mu(z).
\]
Moreover, the normalization constant $N_\mu$ is defined as
\[
N_\mu = \int_0^\infty |f(r)|^2 d\lambda_\mu(r).
\]

If we take the following function
\[
f(r)= r^{-l-\frac{D-1}{2}} u(r)
\]
together with the value $\mu=l+D/2-1$, we find that:
\begin{equation}
\int_0^\infty |f(r)|^2 \ln r\:
d\lambda_\mu(r)
=N_\mu \int_0^\infty \omega (r)\ln r\: dr= N_\mu\langle \ln r\rangle,
\label{eq:f_lnr}
\end{equation}
\begin{equation}
\int_0^\infty |\tilde{f}(p)|^2 \ln p\:
d\lambda_\mu(p)
=N_\mu \int_0^\infty \tilde{\omega}(p)\ln p\: dp= N_\mu\langle \ln p\rangle,
\label{eq:f_lnp}
\end{equation}
and the explicit value of the normalization constant is given by
\begin{equation}
N_\mu=\frac{1}{2^\mu \Gamma(\mu+1)}.
\label{eq:nmu}
\end{equation}
With expressions (\ref{eq:f_lnr}), (\ref{eq:f_lnp}) and (\ref{eq:nmu}) the inequality (\ref{eq:omri_relation}) boils down to the wanted relation (\ref{eq:logarithmic_relation}).

\section{Shannon-entropy-based uncertainty relation for central potentials}
\label{sec:uncertainty_relation}

In this Section we will give and discuss the Shannon-entropy-based uncertainty relation for general $D$-dimensional central potentials, which provides a lower bound to the sum $S[\rho] + S[\gamma]$ in terms of $D$ and the hyperquantum numbers of the corresponding state.

The entropy sum given in Eq. (\ref{eq:entropy_sum_expression}) is bounded from below as
\begin{equation}
S[\rho] + S[\gamma] \ge B_{l,\{\mu\}}
\label{eq:shannon_uncertainty_relation}
\end{equation}
where
\begin{eqnarray}
B_{l,\{\mu\}}&=&2l+D+2\ln\left[\frac{\Gamma\left(l+\frac{D}{2}\right)}{2}\right]-(2l+D-1)\psi\left(l+\frac{D}{2}\right)\nonumber\\
&&+(D-1)\left(\psi\left(\frac{2l+D}{4}\right)
+\ln 2\right)
+2S(\mathcal{Y}_{l,\{\mu\}})
\label{eq:central_bound}
\end{eqnarray}
where $S(\mathcal{Y}_{l,\{\mu\}})$ is given by Eq. (\ref{eq:entropy_y}), and the inequalities (\ref{eq:shannon_uncertainty_w}) and (\ref{eq:logarithmic_relation}) have been taken into account to bound the terms $S[\omega]+S[\tilde{\omega}]$ and $\langle \ln r\rangle + \langle\ln p\rangle$, respectively.
Notice that this bound depends on the hyperangular quantum numbers $(l,\{\mu\})$  and the dimensionality $D$, but not on the principal (energetic) quantum number $n$ because the analytical form of the central potentital $V_D(r)$ was not specified.

\begin{figure}
\begin{center}
\includegraphics[width=14cm]{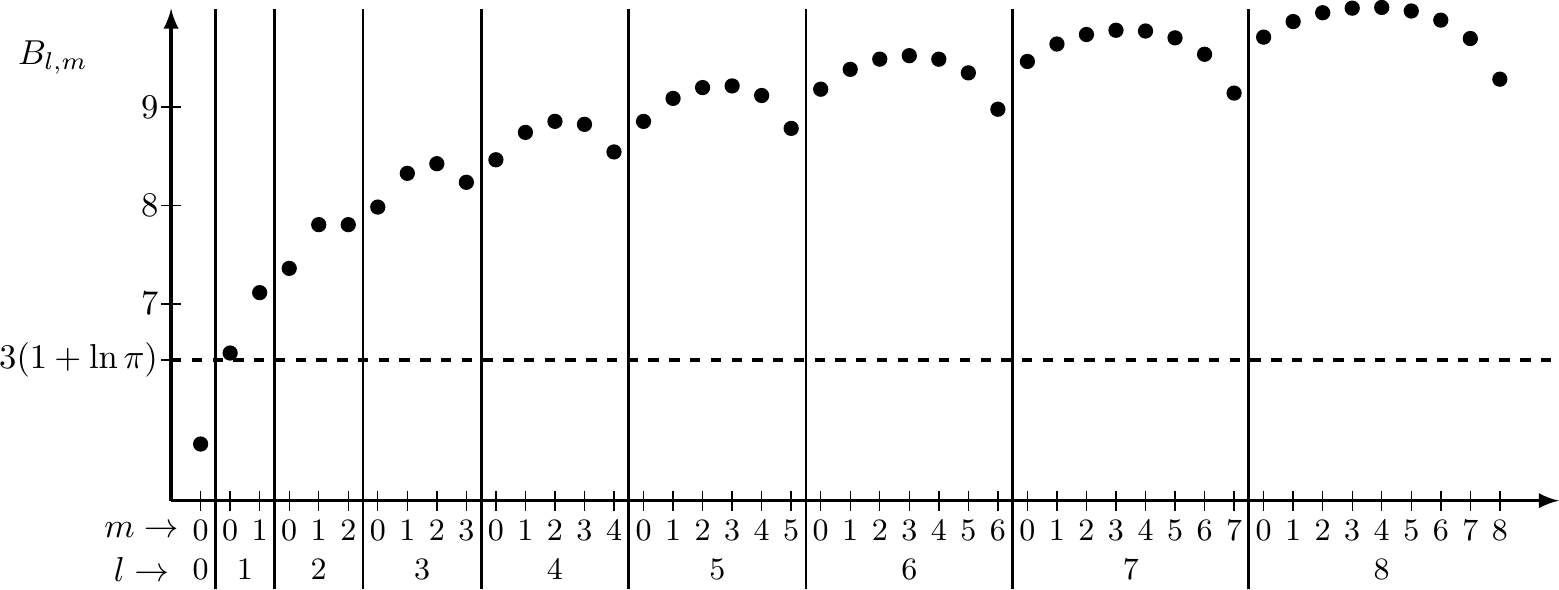}
\end{center}
\caption{General bound given by (\ref{eq:uncertainty_relation_bbm}) (dashed horizontal line) and central bound (\ref{eq:central_bound}) ($\bullet$) for different values of $(l,m)$.}
\label{fig1}
\end{figure}

For completeness, the central bound (\ref{eq:central_bound}) and the general bound given by (\ref{eq:uncertainty_relation_bbm}) are represented in Figure \ref{fig1}, for $D=3$, as a function of the quantum numbers $(l,m)$. The horizontal line represents the general bound (equal to $3(1+\ln\pi)$ in this case), while the points represent the values of the central bound for different quantum numbers $(l,m)$.
The comparison of these bounds
shows that the new lower bound is bigger (so, better) than the general one for all values $(l,m)$ except for $l=m=0$.
For dimensions other than 3, it might occur that the bound does not get improved for more than one set of values $(l,\{\mu\})$; this is e.g., the case of the values $(l,\mu_2,\mu_3)=(0,0,0)$, $(1,0,0)$ and $(1,1,0)$, when $D=4$.
The reason might be due either to the separation done in Eq. (\ref{eq:entropy_sum_expression}) between the sum of the Shannon entropies of the reduced densities and the logarithmic uncertainty sum, or to the fact that, when $l=0$, the logarithmic uncertainty relation (\ref{eq:logarithmic_relation}) does not represent any improvement with respect to the general inequality by Beckner \cite{beckner_pams95}, that is not sharp enough in this case.

Finally, let us point out that in the asymptotic case $l\to\infty$, the Shannon entropy of the hyperspherical harmonics $\mathcal{Y}_{l,\{\mu\}}$ have the asymptotic behaviour \cite{lopezrosa_12}
\[
S(\mathcal{Y}_{l,\{\mu\}}) = O(1) 
\]
for finite values of $\mu_j$, and  
\[
S(\mathcal{Y}_{l,\{\mu\}})= -\frac12 (K-1) \ln l +O(1), 
\]
having the parameters $\mu_j$ ($j=2,\ldots, D-1)$ the values
$l-a_j$, with $a_j\in\mathbb{N}$ fixed, for $j=2,\ldots,K$, being finite otherwise (i.e. for $j=K+1,\ldots,D-1$). Then, the asymptotic behaviour of the new bound (\ref{eq:central_bound}) in these two cases is given by
\[
B_{l,\{\mu\}}=(D-1)\ln l +O(1)
\]
and 
\[
B_{l,\{\mu\}}= (D-K)\ln l + O(1), 
\]
respectively.

Thus, the larger the value of $l$, the bigger is the new central bound, and the larger the improvement with respect to the general bound (\ref{eq:uncertainty_relation_bbm}).

\section{Applications to hydrogenic and isotropic oscillator potentials}
\label{sec:applications}

In this Section we will discuss the Shannon-entropy-based uncertainty relation (\ref{eq:shannon_uncertainty_relation}) for the two main prototypes of central systems: the hydrogenic and isotropic oscillator systems.

\subsection{Hydrogenic systems}

In this case, $V_D(r)=-\frac{1}{r}$ (atomic number $=1$). The densities in the position and momentum space are \cite{yanez_pra94}
\[
\fl \rho_{n,l,\{\mu\}}(\vec{r})=N_{n,l}\,
e^{-\frac{2r}{\eta}} \left(\frac{2r}{\eta}\right)^{2l}\left[L_{n-l-1}^{(2l+D-2)}\left(\frac{2r}{\eta}\right)\right]^2\left|\mathcal{Y}_{l,\{\mu\}}(\Omega_{D-1})\right|^2,
\]
and
\[
\fl \gamma_{n,l,\{\mu\}}(\vec{p})=
\tilde{N}_{n,l}
\frac{(\eta p)^{2l}}{(1+\eta^2 p^2)^{2l+D+1}}\left[C_{n-l-1}^{\left(l+\frac{D-1}{2}\right)}
\left(\frac{1-\eta^2 p^2}{1+\eta^2 p^2}\right)\right]^2
\left|\mathcal{Y}_{l,\{\mu\}}(\Omega_{D-1})\right|^2,
\]
where
\[
N_{n,l}=\left(\frac{2}{\eta}\right)^\frac{D}{2}\frac{(n-l-1)!}{2\eta(n+l+D-3)!},
\]
\[
\tilde{N}_{n,l}=\frac{(n-l-1)!}{2\pi(n+l+D-3)!}4^{2l+D}\Gamma^2\left(l+\frac{D-1}{2}\right)\eta^{D+1},
\]
\[
\eta=n+\frac{D-3}{2},
\]
and $L_k^{(\alpha)}(\cdot)$ and $C_k^{(\alpha)}(\cdot)$ are the Laguerre and Gegenbauer polynomials of degree $k$ and parameter $\alpha$, respectively. The principal quantum number takes the values $n=0,1,2,\ldots$, and $l=0,1,\ldots,n-1$.

\begin{figure}
\begin{center}
\includegraphics[width=14cm]{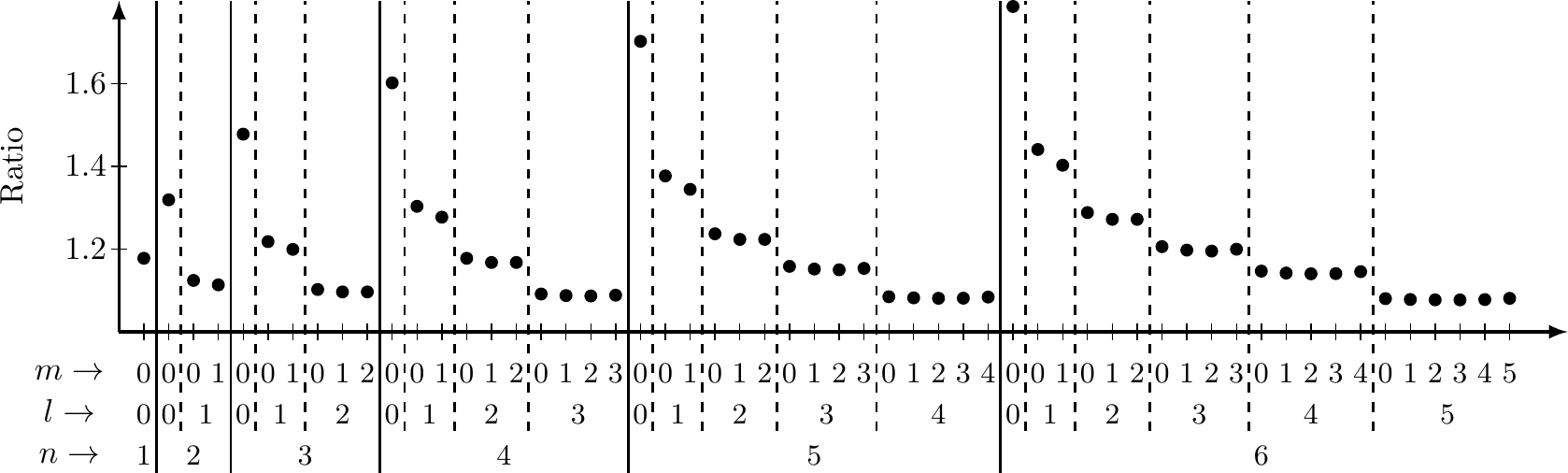}
\end{center}
\caption{Ratio (\ref{eq:hydrogen_ratio}) ($\bullet$) for several states $(n,l,m)$ of the hydrogen atom.}
\label{fig2}
\end{figure}

The Shannon entropies of these two densities have not yet been analytically calculated despite the enormous efforts done since a long time (see Ref. \cite{dehesa_ijqc10} for a recent review). So we have evaluated $S[\rho]$ and $S[\gamma]$ numerically to obtain the entropy sum $S[\rho]+S[\gamma]$. The ratio
\begin{equation}
\Xi_{n,l,m} = \frac{S[\rho]+S[\gamma]}{B_{l,m}}
\label{eq:hydrogen_ratio}
\end{equation}
between the entropy sum and central bound (\ref{eq:central_bound}), for $D=3$, is represented in Figure \ref{fig2} for several states. Naturally, in all the cases $\Xi_{n,l,m}\ge 1$. Moreover, the ratio is lower for larger values of $l$ ($n$ fixed), as suggested by the analysis of Section \ref{sec:uncertainty_relation}.

\subsection{Isotropic oscillator systems}

In this case, $V_D(r)=\frac12\lambda r^2$, $\lambda>0$. The densities in the position and momentum space are \cite{yanez_pra94}
\[
\rho_{n,l,\{\mu\}}(\vec{r})=
\frac{2n! \lambda^{l+\frac{D}{2}}}{\Gamma\left(n+l+\frac{D}{2}\right)}r^{2l}e^{-\lambda r^2} \left[L_n^{\left(l+\frac{D}{2}-1\right)}(\lambda r^2)\right]^2
\left|\mathcal{Y}_{l,\{\mu\}}(\Omega_{D-1})\right|^2,
\]
and
\[
\gamma_{n,l,\{\mu\}}(\vec{p})=
\frac{1}{\lambda^D}\rho_{n,l,\{\mu\}}\left(\frac{\vec{p}}{\lambda}\right),
\]
where
$n,l=0,1,2,\ldots$

\begin{figure}
\begin{center}
\includegraphics[width=14cm]{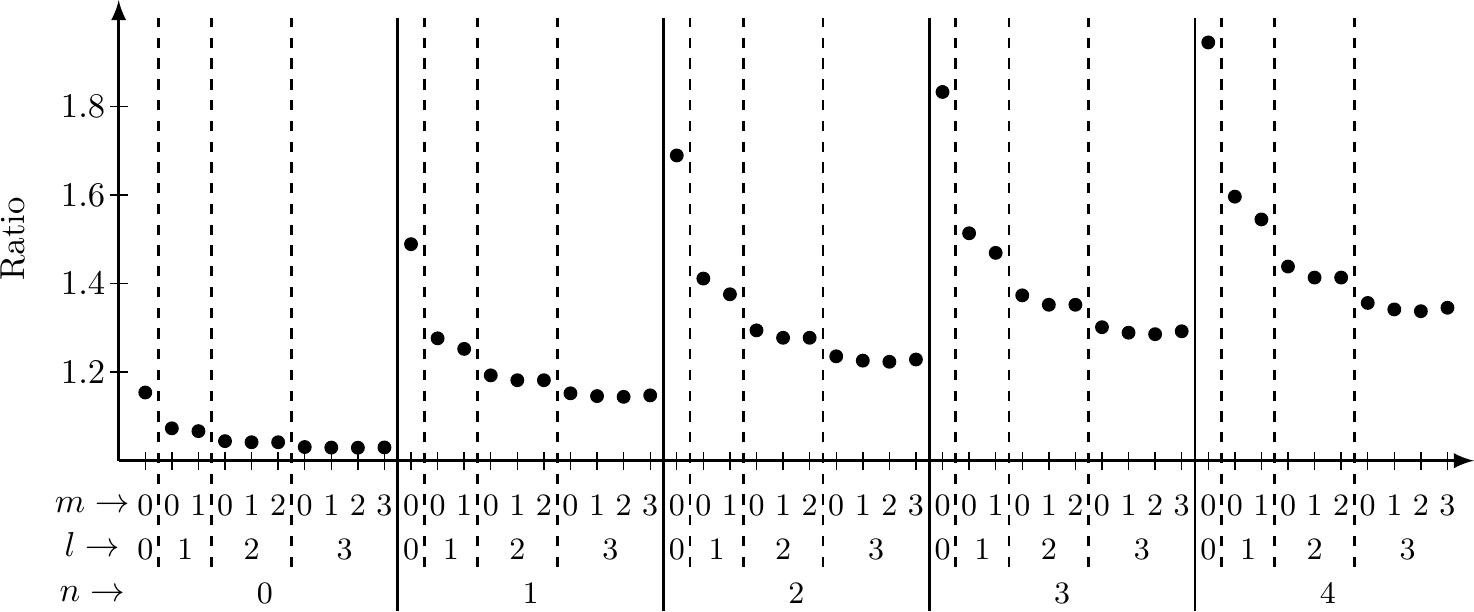}
\end{center}
\caption{Ratio (\ref{eq:ho_ratio}) ($\bullet$) for several states $(n,l,m)$ of the isotropic oscillator.}
\label{fig3}
\end{figure}

Analogously to the previous Subsection, the Shannon entropies, $S[\rho]$ and $S[\gamma]$, for these two densities are evaluated numerically to obtain the entropy sum $S[\rho]+S[\gamma]$. The ratio
\begin{equation}
\Phi_{n,l,m} = \frac{S[\rho]+S[\gamma]}{B_{l,m}}
\label{eq:ho_ratio}
\end{equation}
between the entropy sum and central bound (\ref{eq:central_bound}), for $D=3$ and $\lambda=1$, is represented in Figure \ref{fig3} for several states. Like in the hydrogenic systems, in all the cases $\Phi_{n,l,m}\ge 1$. Moreover, the ratio is lower for larger values of $l$ ($n$ fixed), as suggested by the analysis of Section \ref{sec:uncertainty_relation}. It is worth remarking that for this system the entropy sum of the ground state $(n,l,m)=(0,0,0)$ is equal to $3(1+\ln(\pi))$ (i.e. exactly the bound (\ref{eq:uncertainty_relation_bbm}) for $D=3$). However, as we can see in Figure \ref{fig3}, the ratio of this value with respect to $B_{0,0}$ is $\Phi_{0,0,0}\simeq 1.15$, which is clearly greater than unity obtained with the general bound (\ref{eq:uncertainty_relation_bbm}), as the ground state $(0,0,0)$ of the isotropic oscillator saturates this latter bound. Thus, again we observe that the new bound improves the general bound (\ref{eq:uncertainty_relation_bbm}) for all states $(l,m)$ except when $l=0$.

\section{Conclusions and open problems}
\label{sec:conclusions}

In this work the Shannon-entropy-based uncertainty relation (\ref{eq:uncertainty_relation_bbm}) of the $D$-dimensional quantum systems has been improved for spherically symmetric potentials. We have obtained that the resulting lower bound does not depend on the specific form of the potential since it only depends on the hyperangular quantum numbers  $(l, \{\mu\})$. Moreover we have observed that this bound is indeed a strict improvement of the general bound (\ref{eq:uncertainty_relation_bbm}) for all values of $(l,m)$ except when $l = m = 0$. The latter is because the logarithmic uncertainty relation (\ref{eq:logarithmic_relation}) does not represent for $l =0$ any improvement with respect to the general Beckner inequality (which is not sharp enough for $s$ states). It would be nice to overcome this limitation in the future. Moreover, we have studied and discussed the new ``central'' bound for various states and for some relevant spherically symmetric potentials of Coulomb and oscillator types.

Finally let us point out an important open problem closely related to that resolved here; namely, to improve for central potentials the general R\'enyi-entropy-based uncertainty relation of Bialynicki-Birula, Zozor and Vignat \cite{bialynicki_pra06,zozor:pa07}.

\section*{Acknowledgements}

\L{}R would like to thank all members of the Atomic Physics Group from University of Granada for their kind hospitality. \L{}R acknowledges financial support by the Grant of the Polish Ministry of Science and Higher Education for years 2010-2012.
JSD and PSM are very grateful for partial support to Junta de Andaluc\'{\i}a
(under grants FQM-4643 and FQM-2445) and Ministerio de Ciencia e Innovaci\'on under project FIS2011-24540. JSD and PSM belong to the Andalusian research group FQM-0207. \L{}R and PSM are grateful for the GENIL-PYR-2010-27  research grant.

\bibliographystyle{unsrt}
\bibliography{uncertainty_relation}

\end{document}